\documentclass{article}
\usepackage{LaThuileFPSpro}
\usepackage{graphicx,amssymb}
\newcommand{\0} {0\hspace{-1.15ex}/}
\begin{document}
\title{ 
  QCD PHYSICS AT THE TEVATRON
  }
\author{
  Giuseppe Latino \\
  (for the CDF and D$\0$ Collaborations) \\
  {\em University of New Mexico, 800 Yale Blvd. NE Albuquerque, NM 87131, USA  } \\
  {\em e-mail: latino@fnal.gov}
  }
\maketitle

\baselineskip=11.6pt

\begin{abstract}
  Results on recent QCD measurements performed at the Tevatron $p\bar p$ 
  Collider at $\sqrt{s}$ = 1.96 TeV are here reported. The inclusive jet 
  and dijet mass cross sections are compared to NLO pQCD calculations and 
  to Run I results. The production rates and kinematic properties of W + jets 
  production processes are compared to ``enhanced'' LO theoretical 
  predictions. Non-perturbative ``soft'' interactions leading to  
  the underlying event are studied and compared to QCD Monte Carlo 
  phenomenological models.   
\end{abstract}
\newpage
\section{Introduction}
Measurements aimed to test the predictions of Quantum Chromodynamics (QCD), the currently 
accepted theory of the strong interactions among quarks and gluons, represent a very 
important part of the physics program carried out at the Tevatron $p\bar p$ Collider.
The large amount of data expected to be accumulated during Run II and the increase in 
center-of-mass energy ($\sqrt{s}$) from 1.8 to 1.96 TeV, 
give CDF\cite{CDF-II} and D$\0$\cite{D0-II} experiments an unique opportunity to make 
precision tests of next-to-leading order perturbative QCD (NLO pQCD) and, by looking for 
deviations from theory, to search for new particles and new interactions down to a 
distance scale of $\sim10^{-19}$ m.

An optimal understanding of QCD in hadron collisions allows to improve the constraints 
on the fundamental parameters of the theory, ${\alpha}_s$ 
and the parton distribution functions (PDFs); results in a better control on the standard 
QCD production which represents the main background for most of the processes of interest, 
such as top and Higgs production; gives phenomenological input for the modeling of the 
non-perturbative regime (where the theory fails in its predictivity), such as the 
``soft'' interactions generating the underlying event which accompanies the ``hard'' 
collision. 
\section{Inclusive Jet Cross Section}
\subsection{Experience from Run I}
\begin{figure}[!htb]
  \vspace{6.1cm}
  \includegraphics{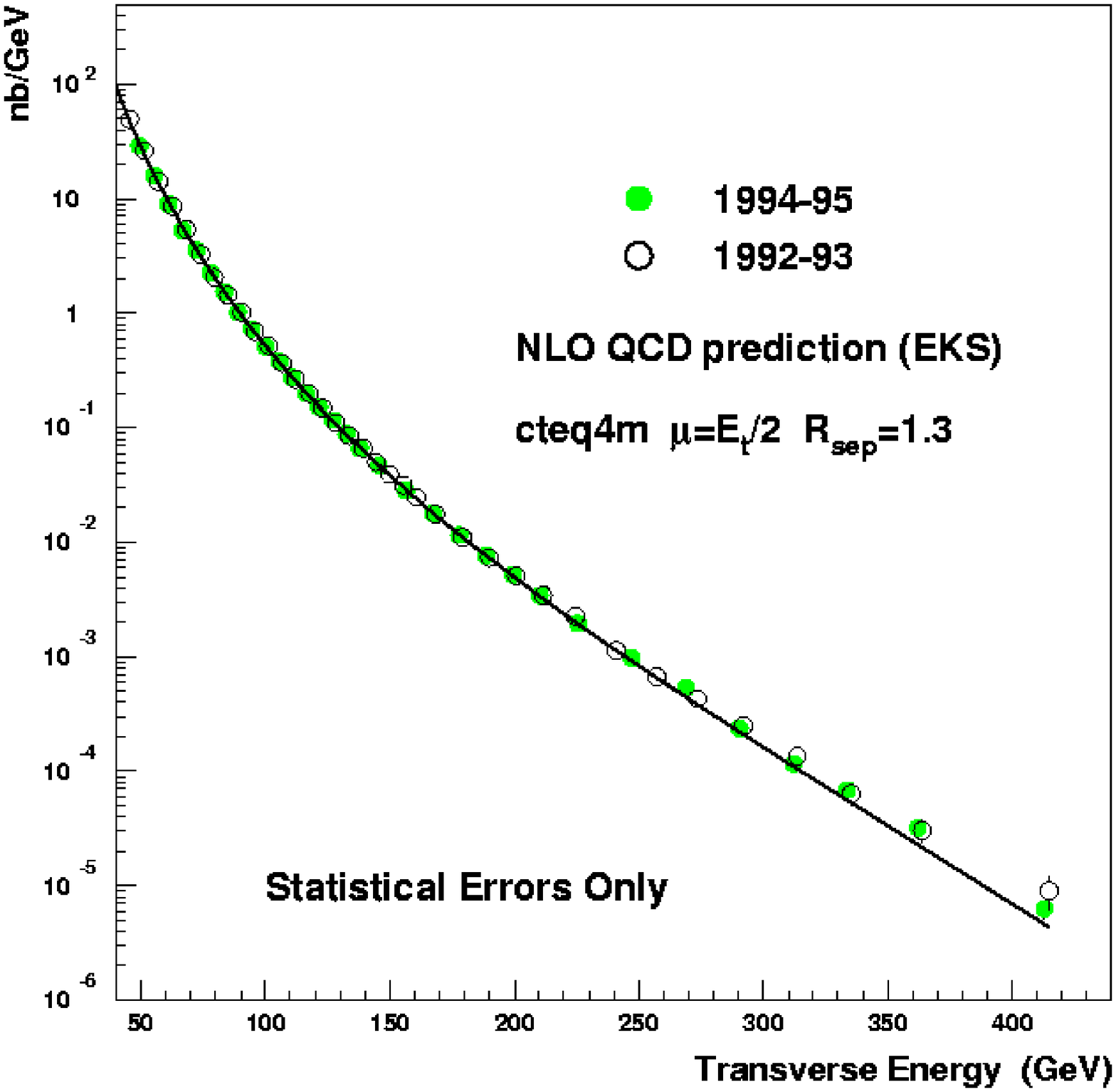}
  \includegraphics{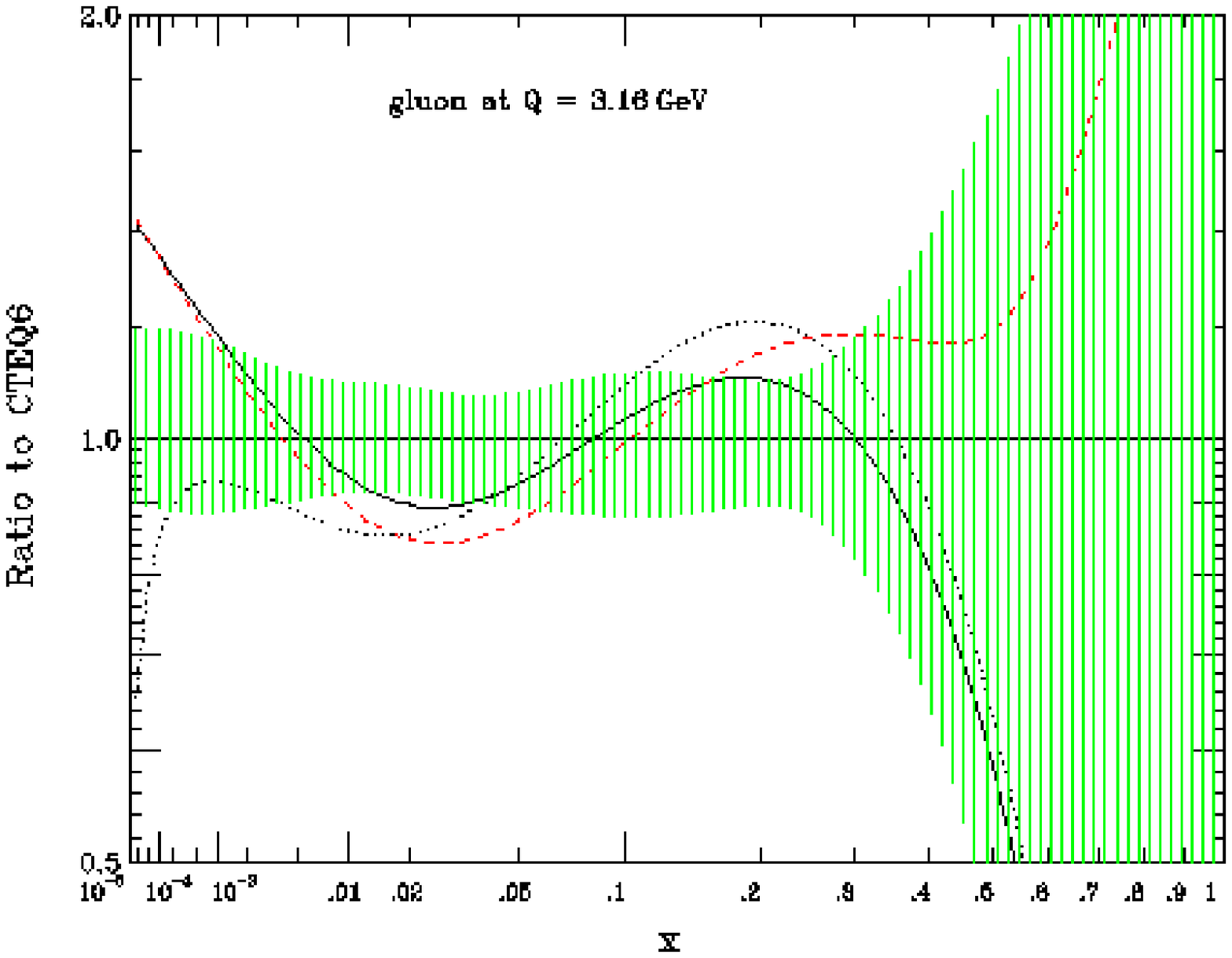}
  \caption{\it
    Left: inclusive jet cross section as measured in CDF
    using 87 $pb^{-1}$ of Run I data at $\sqrt{s}$ = 1.8 TeV\cite{CDF-new-incl-jet}.
    Right: uncertainty band for the gluon distribution function at $Q^2$ = 10 ${GeV}^2$ in
    the CTEQ6M set\cite{CTEQ6}; comparisons to CTEQ5M1 (solid), CTEQ5HJ (dashed) and 
    MRST01\cite{MRST01} (dotted) are also shown in terms of ratios to CTEQ6M.
    \label{RunI_CDF_JetXsec} }
\end{figure}
During Run I, the CDF and D$\0$ Collaborations performed several QCD measurements which, 
in general, were found to be in reasonable agreement with theoretical expectations. 
However, initial inclusive jet cross section
measurements showed an excess of data at high ${E_T}^J$ over NLO pQCD predictions, 
which raised great interest among the high energy physics community and stimulated a 
reevaluation of the uncertainties associated to theoretical calculations\cite{CDF-old-incl-jet}. 
Subsequent studies have demonstrated that such excess can be explained within the 
Standard Model in terms of a larger than expected gluon distribution at high $x$. 
A better agreement of data versus theory was actually observed in subsequent 
measurements involving an increased data sample when PDFs with an enhanced 
gluon contribution at high $x$ (CTEQ4HJ) were considered\cite{CDF-new-incl-jet,D0-new-incl-jet} 
(see fig.\ref{RunI_CDF_JetXsec}, left). Given the large uncertainty in the NLO 
calculations arising from the flexibility allowed by current knowledge of PDFs as well as  
the uncertainty in the experimental results, CDF and D$\0$ data were found to be 
consistent between them, with previous measurements and with NLO pQCD. 

Most recent global PDFs fits have used results from these Run I analyses so to include 
the Tevatron high $E_T$ jet data in the determination of the high $x$ gluon distribution. 
In particular, by involving jets in a range of
rapidity intervals up to the forward region, the D$\0$ measurement allowed to constrain the 
partons over a much wider $x$ kinematical range.
These new PDF sets (CTEQ6\cite{CTEQ6} and MRST02\cite{MRST02})
represent the most complete information available for Run II QCD predictions,
but are still characterized by a big uncertainty on the gluon distribution at 
high $x$ (see fig.\ref{RunI_CDF_JetXsec}, right). 

With larger data samples collected with a higher cross section\footnote
{
The increase in $\sqrt{s}$ from 1.8 to 1.96 TeV leads to a
significant increase in jet cross section at high ${P_T}^J$, about a factor of 5
at ${P_T}^J$ $\sim$ 600 GeV. 
},
as well as with the improved performances of the upgraded CDF and D$\0$ detectors,
Run II jet measurements are expected to reduce this uncertainty so to give the best 
constraint on PDFs before the LHC era.
\subsection{Preliminary Run II measurements}
Both CDF and D$\0$ experiments have recently performed preliminary Run II measurements of the 
inclusive jet cross section.
\begin{figure}[!htb]
  \vspace{5.2cm}
  \includegraphics{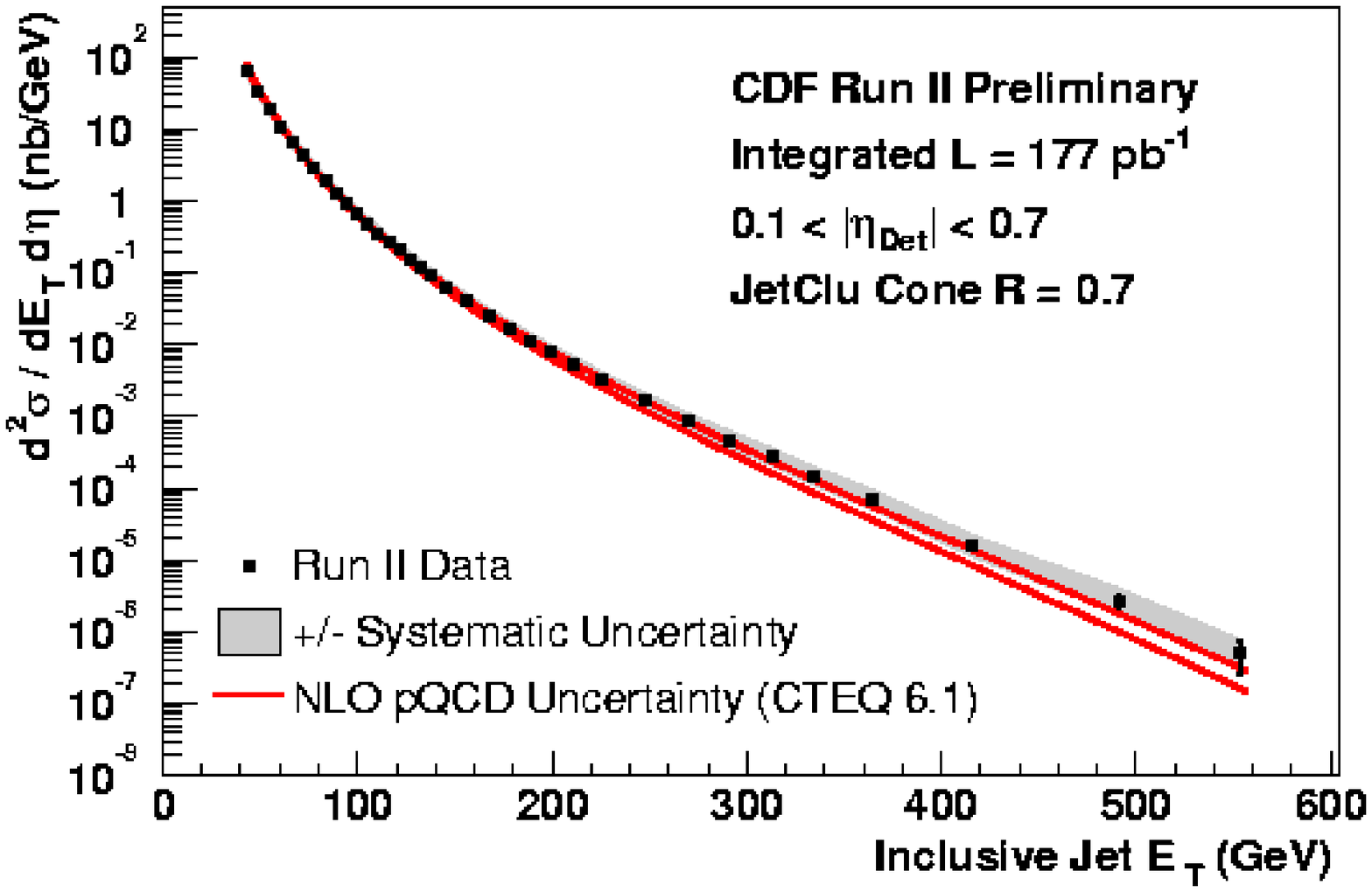}
   \includegraphics{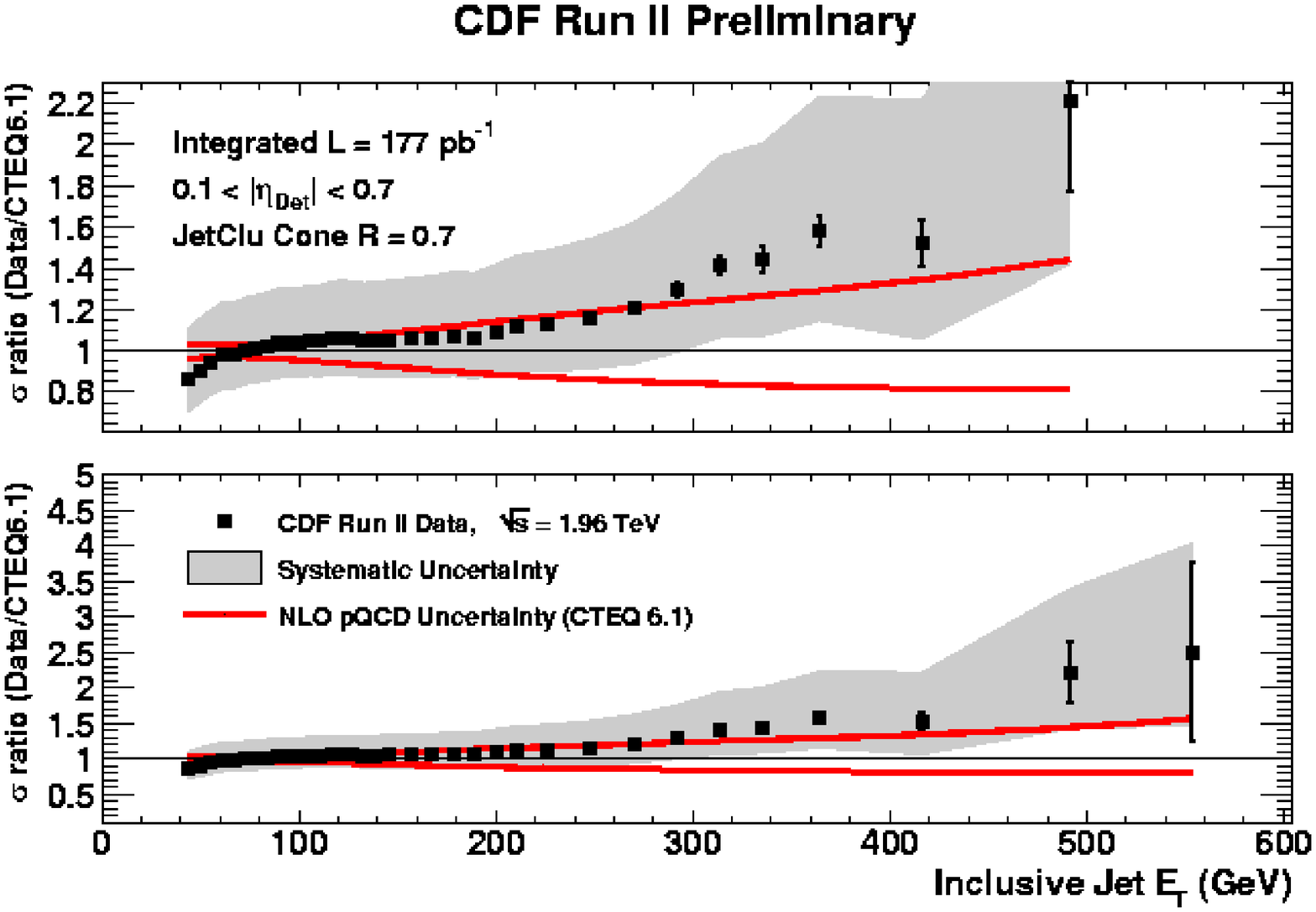}
  \vspace{5.4cm}
  \includegraphics{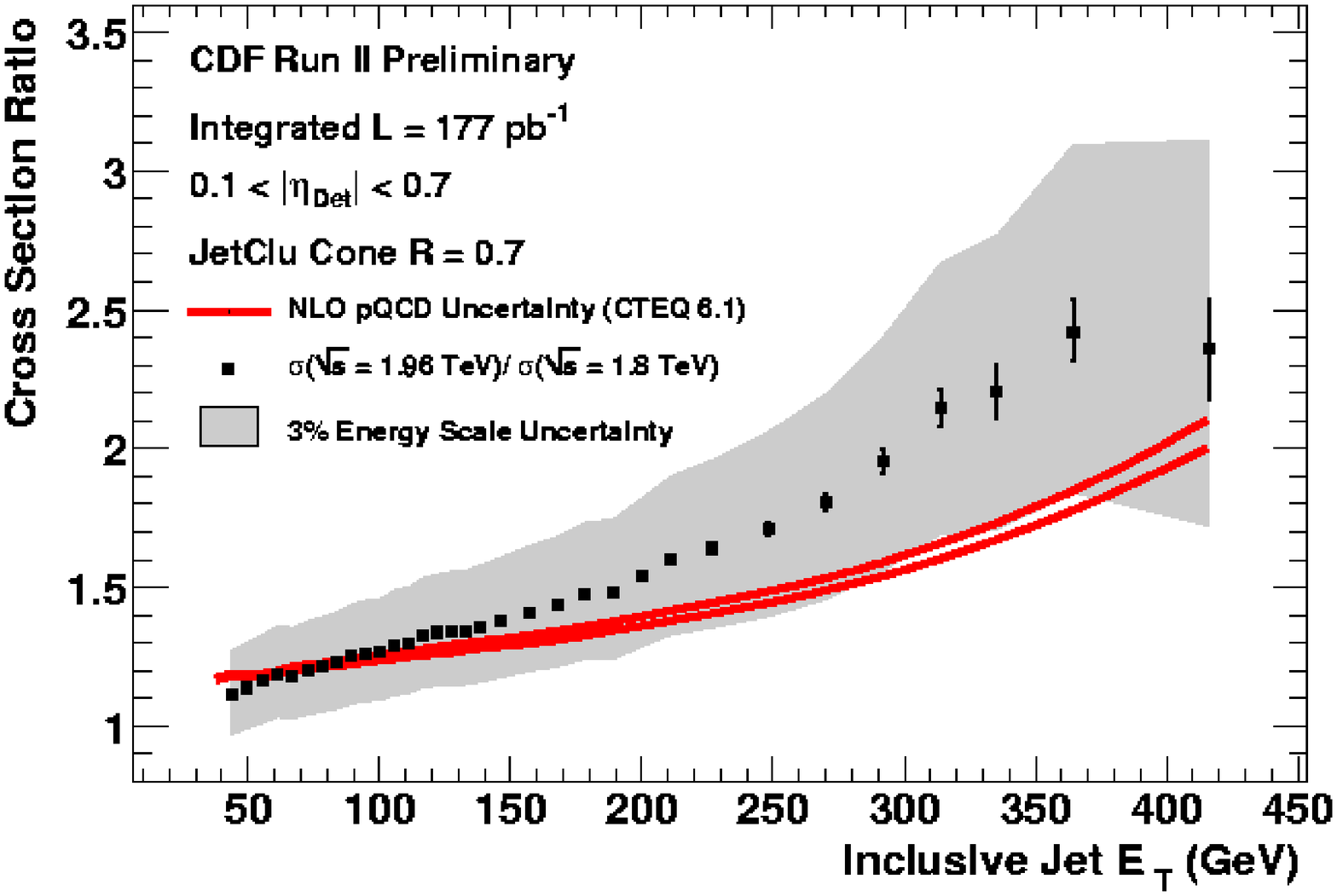}
  \caption{\it
    Inclusive jet cross section as measured in CDF analyzing 177 $pb^{-1}$ of Run II
    data at $\sqrt{s}$ = 1.96 TeV.
    Upper left: the results are compared on a log scale to the NLO pQCD
    prediction from EKS ($\mu$ = $E_{T}^{J}/2$, $R_{sep}$ = 1.3, CTEQ6.1M PDFs).
    Upper right: ratio of data to theoretical
    prediction on a linear scale (the upper plot is a zoom obtained by excluding the last bin).
    Lower: Ratio of Run II inclusive jet cross section to Run I one compared to theoretical
    prediction. The increase in cross section due to the increase in $\sqrt{s}$ is
    evident.
    Data points include the statistical error, the error band represents the change in the
    cross section due to the 3$\%$ energy scale shift corresponding to the dominant
    experimental uncertainty. The two solid lines represent the dominant theoretical
    uncertainty due to PDFs.
   \label{RunII_CDF_JetXsec} }
\end{figure}

In CDF, jets were reconstructed using the same Run I cone algorithm 
(JetClu\cite{jetclu}, $R$ = 0.7) in the rapidity range 0.1 $<$ $|\eta|$ $<$ 0.7; 
the same Run I correction procedures were also used to properly 
correct the jet spectrum for both detector and physics 
effects affecting jet measurements\cite{CDF-new-incl-jet}.
Four different samples (from 20, 50, 70 and 100 GeV jet $E_T$ trigger threshold 
requirements), were combined after correcting for trigger prescale and efficiency 
effects giving a jet $E_T$ spectrum already extending the Run I reach by about 150 GeV
and spanning approximately 9 orders of magnitude 
\footnote{
${E_T}^J$ = 666 GeV is the highest jet $E_T$ so far observed.
}.
The results reported here and summarized in fig.\ref{RunII_CDF_JetXsec} correspond to an 
integrated luminosity of 177 $pb^{-1}$ collected during the period 2002-3. In general, 
a reasonably good agreement is found within errors between
data and NLO pQCD theory (EKS\cite{EKS}) using the latest CTEQ PDF sets\cite{CTEQ6}.
Systematic uncertainties are found to be dominanted by the error on jet energy scale in
data and on high $x$ gluon distribution function in theoretical calculations.
Work is in progress in order to reduce the energy scale uncertainty, to use other jet 
algorithms like MidPoint and Kt\cite{JetAlg} and to extend the analysis to forward jets. 
\begin{figure}[!htb]
  \vspace{5.2cm}
 \includegraphics{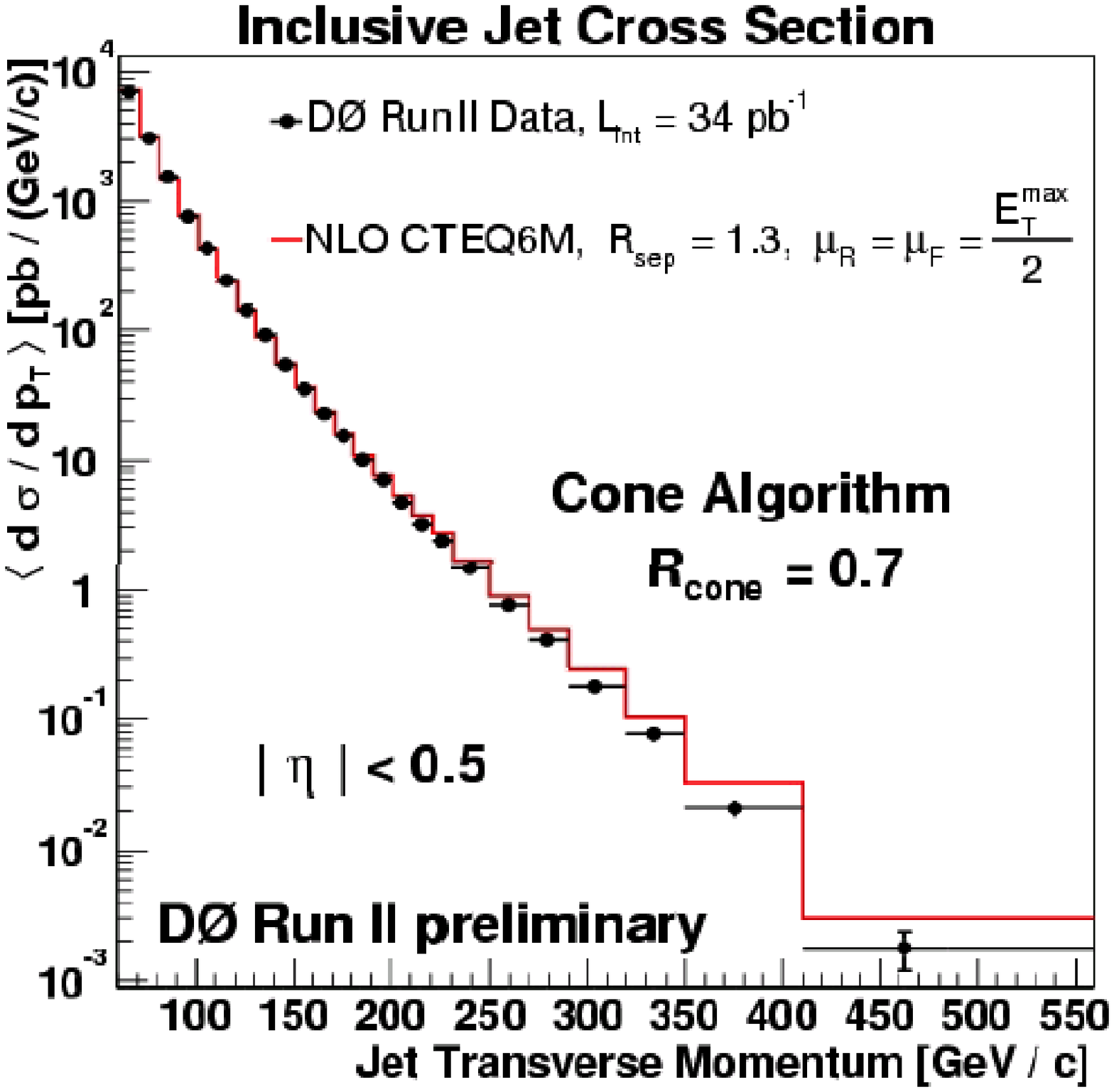}
  \vspace{5.0cm}
 \includegraphics{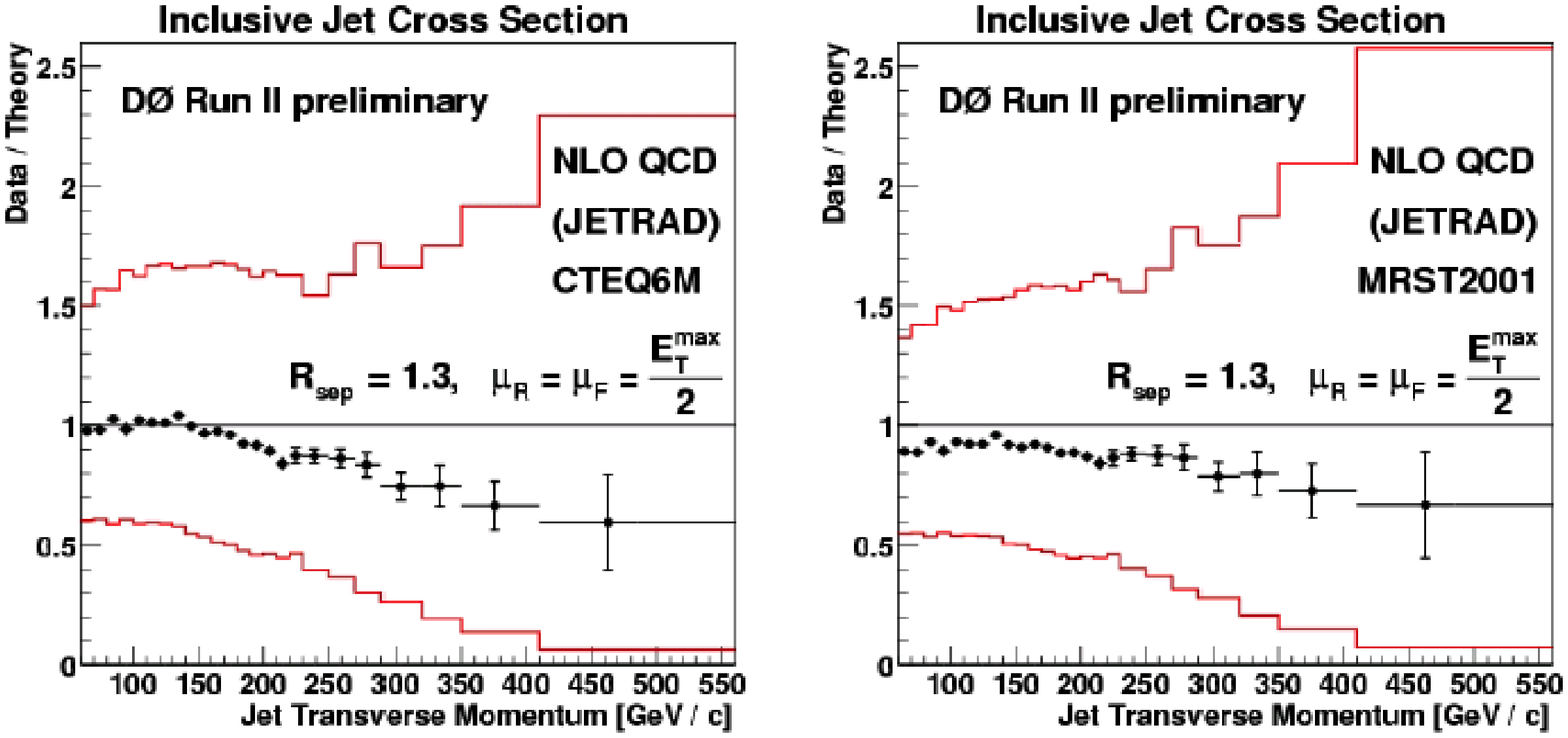}
  \caption{\it
    Inclusive jet cross section as measured in D$\0$ analyzing 34 $pb^{-1}$ of Run II
    data at $\sqrt{s}$ = 1.96 TeV.
    Upper: the results are compared on a log scale to the NLO pQCD
    prediction from JETRAD ($\mu$ = $P_{T}^{max}/2$, $R_{sep}$ = 1.3, CTEQ6M PDFs).
    Lower: linear comparison of data to theoretical prediction with CTEQ6M (left) and
    MRST01 (right) PDF sets.
    Data points include the statistical error, the error band represents the total systematic
    uncertainty.
   \label{RunII_D0_JetXsec} }
\end{figure}

A completely new analysis was performed in D$\0$.
Jets were reconstructed with an ``optimized'' cone algorithm
(MidPoint\cite{JetAlg}, $R$ = 0.7) and then corrected with a new set of jet corrections as
derived from $\gamma$-jet and dijet balancing studies on data\cite{d0_jet_calib} as well as 
from Monte Carlo simulations. The jet energy scale derived with this method was 
characterized by large statistical and systematic uncertainties increasing with jet energy 
due to extrapolation (mostly generated by small $\gamma$+jet sample statistics above 200 GeV). 
Jet measurements were restricted to the central region ($|\eta|$ 
$<$ 0.5) to limit the impact of these uncertainties which are the dominant ones, but are 
expected to be gradually reduced with an increasing amount of collected data. 
Finally, jet resolution effects were taken into account by unsmearing the measured jet 
spectrum with an unfolding procedure.
Fig.\ref{RunII_D0_JetXsec} summarizes the results of a preliminary measurement 
from the analysis of four different jets samples 
(from 25, 45, 65 and 95 GeV jet $P_T$ trigger thresholds) based on an 
integrated luminosity of 34 $pb^{-1}$. An agreement within the rather large uncertainties 
(dominated by the systematic error on the jet energy scale) is 
observed between data and NLO theory (JETRAD\cite{JETRAD}) using CTEQ6 and MRST01 
PDF sets. 
\section{Dijet Invariant Mass Distribution}
\begin{figure}[!htb]
  \vspace{4.4cm}
  \includegraphics{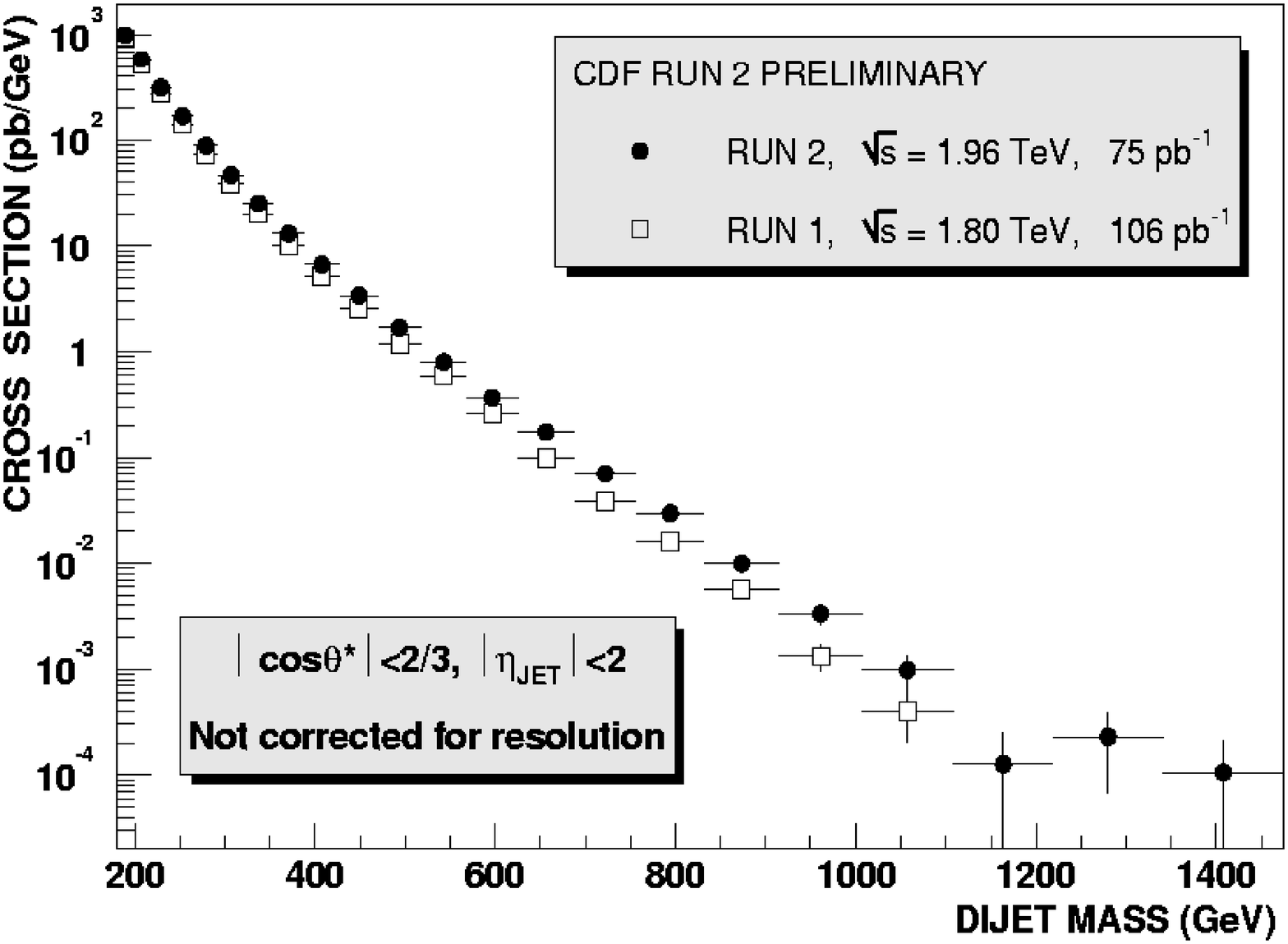}
  \includegraphics{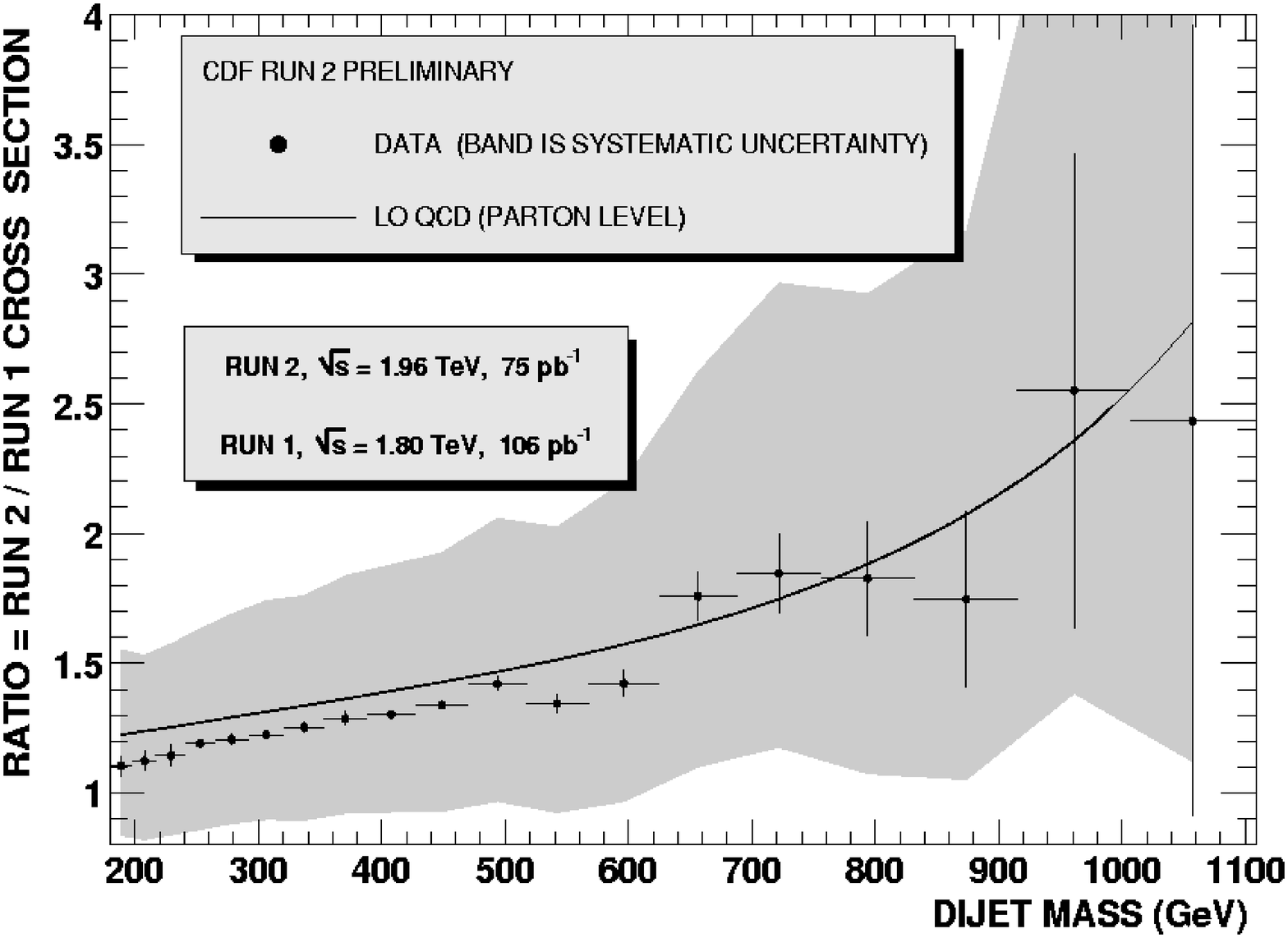}
  \vspace{4.6cm}
 \includegraphics{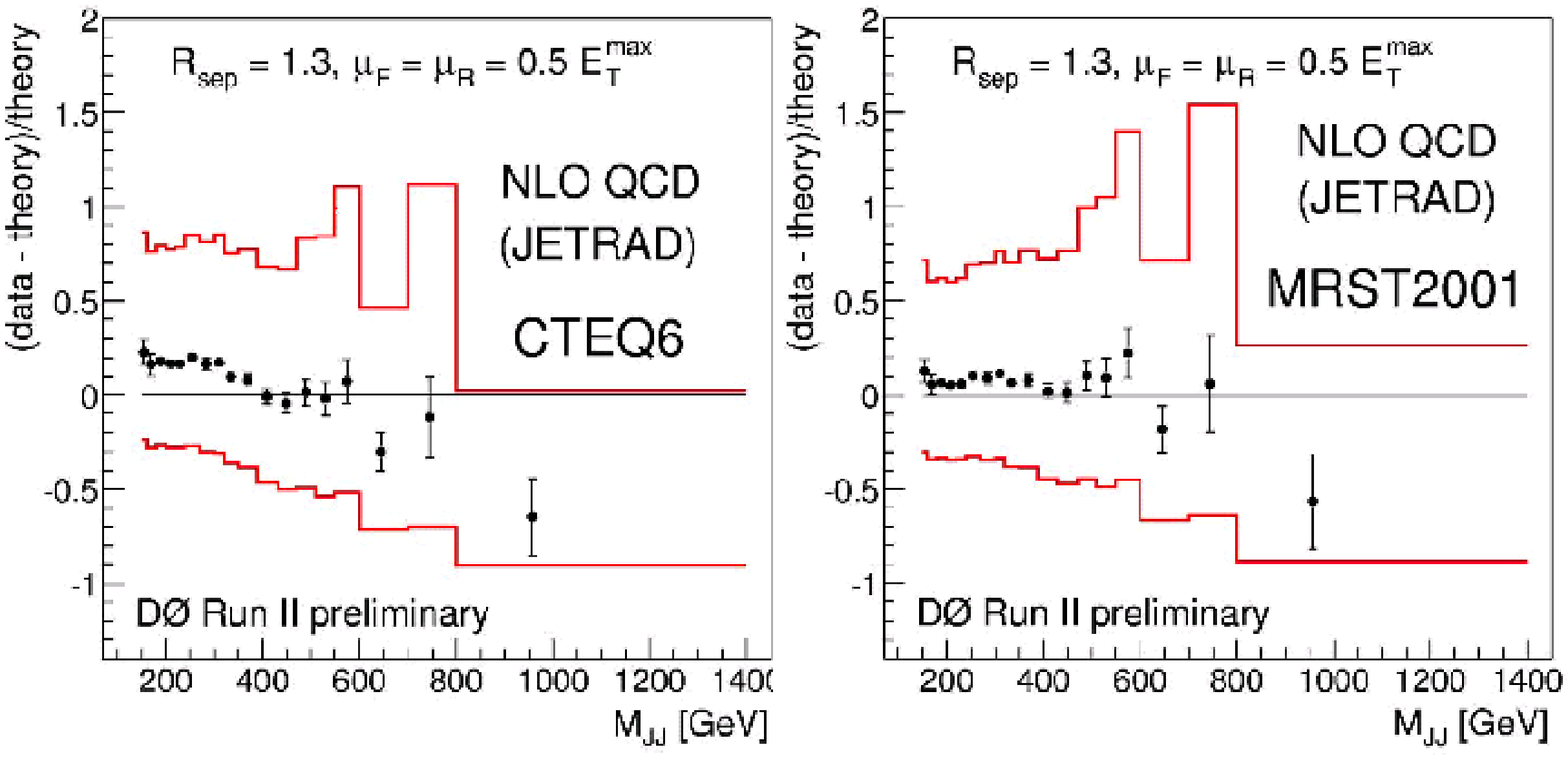}
  \caption{\it
    Upper: comparison of the dijet mass spectrum as measured in CDF using 106 $pb^{-1}$
    and 75 $pb^{-1}$ of Run I and Run II data respectively (left);
    the CDF Run II over Run I dijet mass cross section ratio is compared to a LO QCD
    calculation (right), error bars on data points represent the statistical error,
    the shaded area represents the systematic uncertainty.
    Lower: percentage difference between data and theory relative to theory for the
    dijet mass cross section measured in D$\0$ using 34 $pb^{-1}$ of Run II data;
    the NLO theoretical calculation (JETRAD, $\mu$ = $P_{T}^{max}/2$,
    $R_{sep}$ = 1.3) involves the CTEQ6M (left) and
    MRST01 (right) PDF sets. Data points include the statistical error,
    the solid band represents the total systematic uncertainty.
    \label{RunII_CDF-D0_DiJetMassXsec} }
\end{figure}

The dijet mass cross section measurement represents another powerful test of NLO pQCD.
It is also sensitive to new physics such as quark compositeness (expected to give departure 
from theory at high masses) and new models beyond the Standard Model predicting new particles 
decaying to dijet (expected to generate a peak in the dijet invariant mass 
spectrum at high values).

The CDF and D$\0$ Collaborations have both performed a preliminary Run II measurement 
of the dijet mass spectrum using similar criteria as well as the same jet trigger 
samples as the ones involved for the inclusive jet cross section study. 
In particular, considering the two leading jets per event in order to reconstruct the 
invariant mass observable, the same Run I cone algorithm and analysis strategies 
(Jetclu R = 0.7 jets reconstructed up to $|\eta|$ $<$ 2.0 and corrected a the parton level) 
were used in CDF, while a completely new analysis was performed in D$\0$ using the 
MidPoint algorithm and applying the same jet scale corrections and unsmearing procedures 
as the ones used for the inclusive jet cross section measurement to jets reconstructed 
in the same rapidity range ($|\eta|$ $<$ 0.5).

Results from CDF, corresponding to an integrated luminosity of 75
$pb^{-1}$, are summarized in the upper plots of fig.\ref{RunII_CDF-D0_DiJetMassXsec}.  
The comparison to Run I data (upper left) shows that the Run II dijet mass spectrum already
extends the Run I one by about 350 GeV
\footnote{
The highest dijet mass value was $M_{JJ}$ = 1364 GeV.
}.
The effect of the increase in cross section as consequence of the increase in $\sqrt{s}$ is more
evident in terms of the Run II over Run I cross section ratio (upper right). A good agreement 
with a LO calculation is observed within systematic uncertainties (dominated by the one on 
jet energy scale). 
The ongoing work performed in order to optimize the jet corrections, to use the MidPoint 
algorithm as well as to implement a comparison to a NLO calculation is expected to update these 
preliminary results.

The lower plots of fig.\ref{RunII_CDF-D0_DiJetMassXsec} summarize the results
obtained from D$\0$ relatively to an integrated luminosity of 34 $pb^{-1}$. Here the dijet mass
spectrum measurement is compared to a NLO calculation (JETRAD) on a linear scale 
considering two different choices for the PDF sets: CTEQ6 (lower left) and 
MRST01 (lower right). As in the case of the inclusive jet cross section, a reasonable 
agreement between data and NLO theory is observed within the errors (dominated by the jet 
energy uncertainty) for both the PDF sets. 
\section{W + Jets Production}
The production of W bosons in association with high energy hadronic jets at the Tevatron 
$p\bar p$ collider provides the opportunity to test pQCD at large momentum transfer. 
Understanding the QCD production for W + jets events is also important 
for the study of many Standard Model and new physics processes 
(such as top quark measurements and Higgs and Susy searches) for which it represents 
a major background. 
\begin{figure}[!htb]
  \vspace{4.9cm}
 \includegraphics{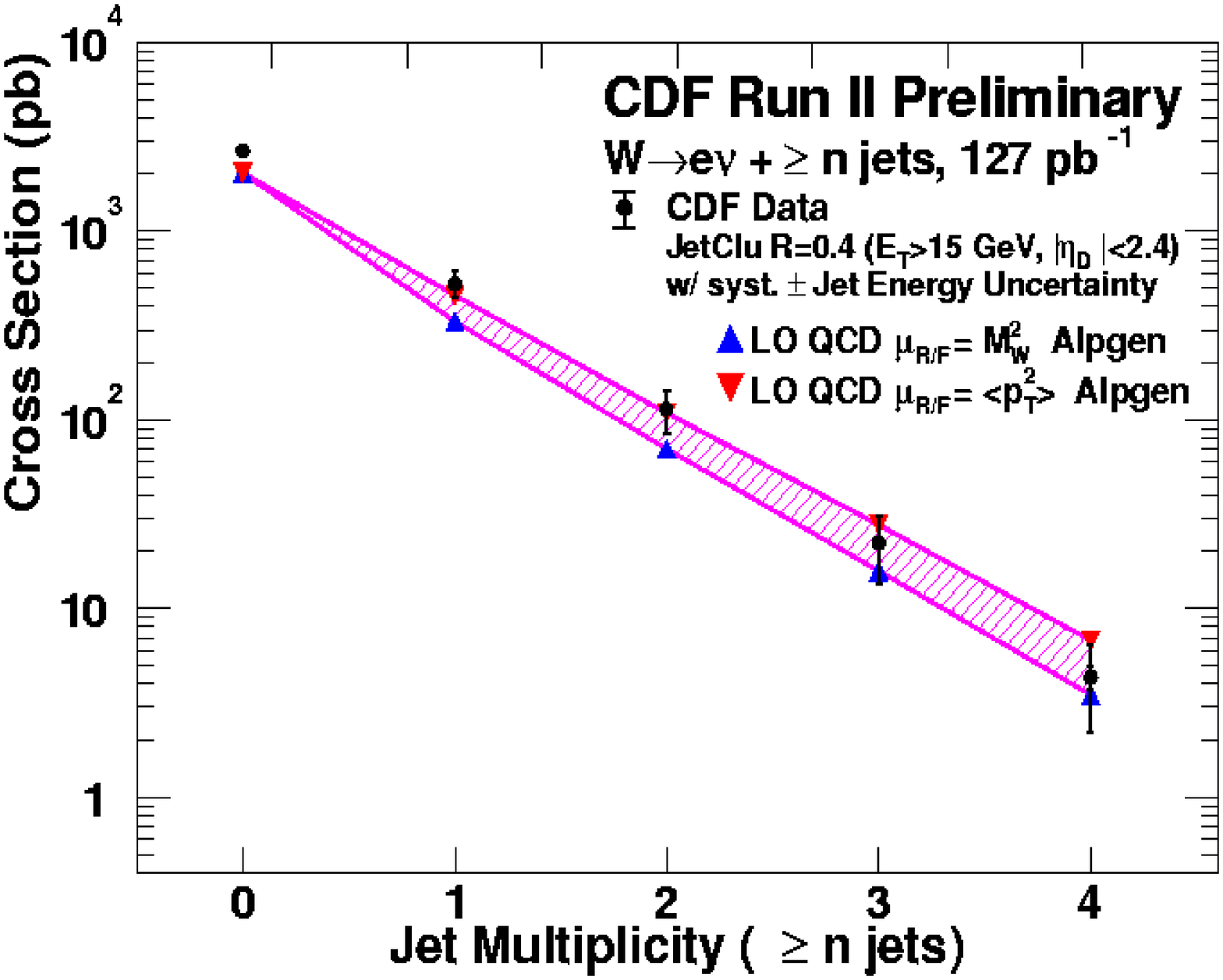}
 \includegraphics{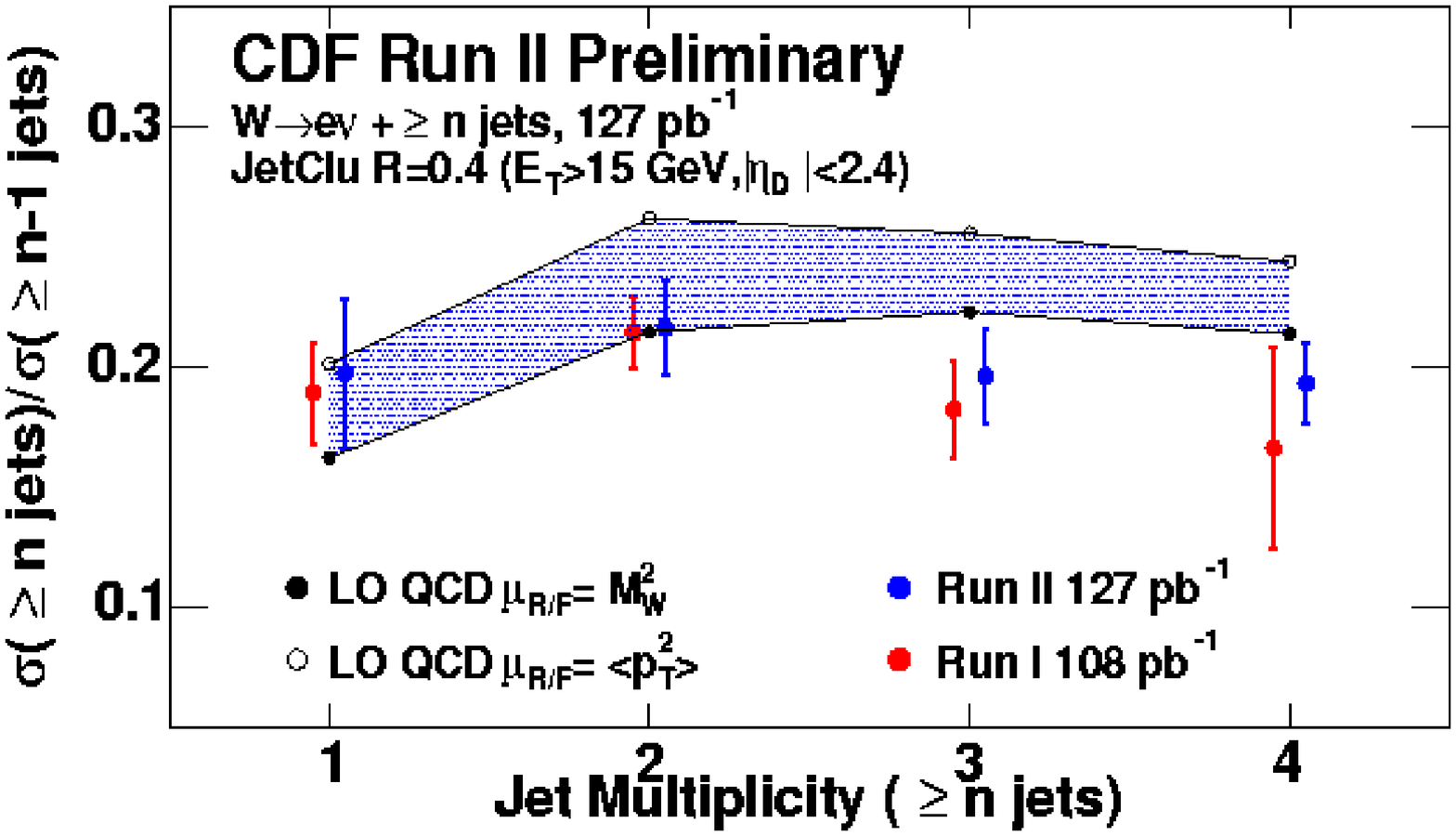}
  \vspace{4.7cm}
 \includegraphics{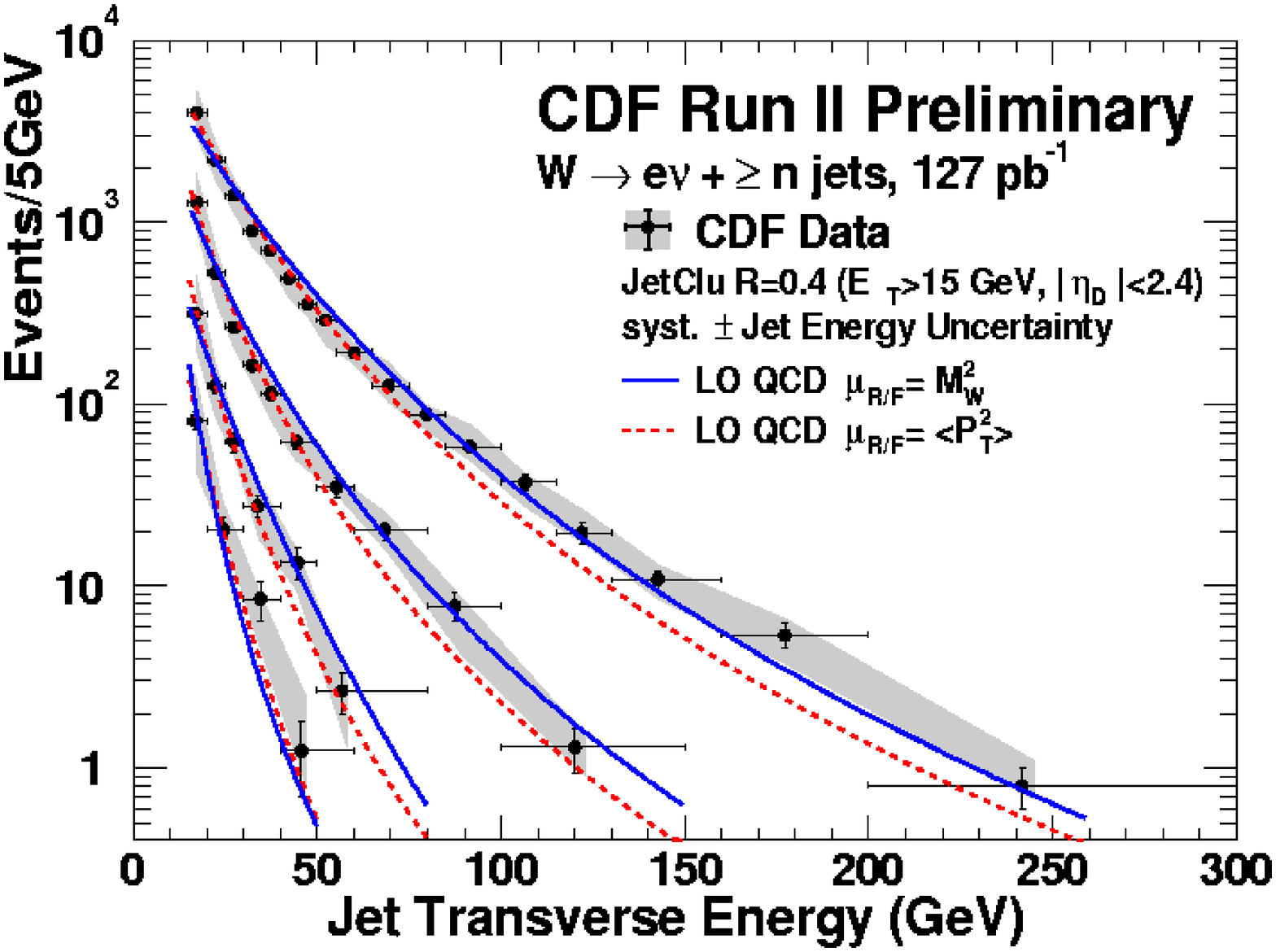}
 \includegraphics{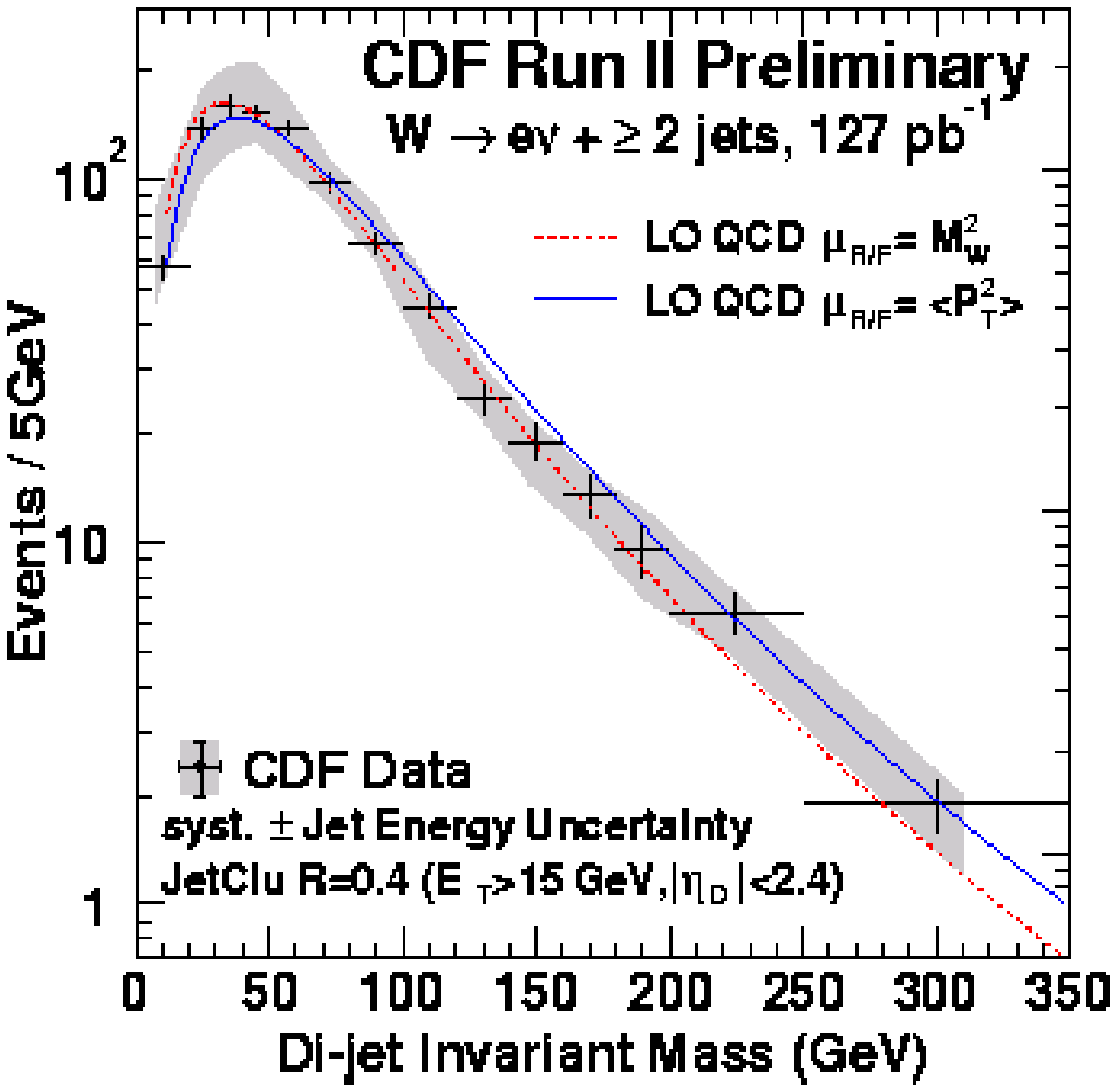}
  \caption{\it
    Upper: W + $\ge$ n jets cross sections measured in CDF analyzing 127 $pb^{-1}$ of Run II
    data at $\sqrt{s}$ = 1.96 TeV compared to LO theoretical predictions (left); ratio of
    cross sections (${\sigma}_{W\ge n jets}$/${\sigma}_{W\ge n-1 jets}$) as a function of
    jet multiplicity (right). Error bars on data points include both statistical and
    systematic uncertainties. The filled band corresponds to the variation of the
    theoretical predictions with ${\mu}_{R/F}$.
    Lower: differential cross section as a function of jet $E_T$ for the leading jet in
    W $\ge$ 1 jets events, the second highest jet $E_T$ in W $\ge$ 2 jets events and so on
    up to W $\ge$ 4 jets events, respectively from top to bottom (left); $M_{JJ}$
    distribution for the two leading jets in W $\ge$ 2 jets events (right).
    Data points include statistical errors, while the band represents the jet energy
    systematic uncertainty. The two lines are fits to the theoretical distributions
    for two different values of ${\mu}_{R/F}$.
   \label{RunII_CDF_WJETS} }
\end{figure}

A preliminary study was performed in CDF using 127 $pb^{-1}$ of Run II data at 
$\sqrt{s}$ = 1.96 TeV collected in the period 2002-3. The production rates and 
kinematic properties of W + $\ge$ n jets events were studied and compared to the predictions
of an ``enhanced'' LO QCD Monte Carlo (MC). The well understood electroweak W $\rightarrow$ 
e$\nu$ decay channel was considered as it provides an efficient and clean identification 
of W candidate with a low background contamination. Jets, reconstructed with the standard 
CDF cone algorithm (JetClu) with R = 0.4\footnote
{The selection of this cone size was motivated by related studies on top quark physics.
} 
and corrected to the parton level, were selected 
by requiring ${E_T}^J$ $>$ 15 GeV and $|\eta|$ $<$ 2.4. QCD production, faking W + n jet events, 
was found to be the largest source of background in all jet multiplicity bins up to n = 3, 
with top production becoming dominant at higher ($\ge$ 4) multiplicities.
The inclusive cross sections for $p\bar p$ $\rightarrow$ ($W^{\pm}$ $\rightarrow$ 
$e^{\pm}\nu$) + n jets production (${\sigma}_{W\ge n jets}$) were obtained after 
backgrounds and detection efficiencies (as derived from data and MC studies) were properly 
taken into account.
The ``enhanced'' LO QCD MC consisted of the ALPGEN\cite{alpgen} program ($\mu$ = $M_{W}^2$, 
$<$$(P_{T}^J)^2$$>$; CTEQ5L PDFs), used to generate W $\rightarrow$ e$\nu$ + n partons 
(n = 1 to 4) at LO, interfaced with HERWIG\cite{herwig} for the implementation of the shower 
evolution for the initial and final state radiation, of the hadronization process
and for the simulation of the undelying event; finally, the full detector simulation and
event reconstruction were applied.

The measured W + $\ge$ n jets cross section as a function of the jet multiplicity is 
compared to the theoretical calculations in fig.\ref{RunII_CDF_WJETS} (upper left). 
Error bars on data points represent both statistical and systematic 
uncertainties which, being dominated by the systematic error on jet energy and ranging from 
$\sim$ 13$\%$ for ${\sigma}_{W\ge 1 jets}$ to $\sim$ 45$\%$ for ${\sigma}_{W\ge 4 jets}$, 
clearly limit the sensitivity of this measurement. 
The filled band shows the variation of the theoretical prediction with the 
renormalization/factorization scale (${\mu}_{R/F}$), the dominant theoretical systematic 
uncertainty as expected for a LO calculation\footnote{
The inclusive W cross section (${\sigma}_{W\ge 0 jets}$) was generated at LO by HERWIG.  
This LO calculation, being independent of QCD effects, is not sensitive to the choice for 
${\mu}_{R/F}$ and is lower than data due to the LO approximation.
}.
Also shown in fig.\ref{RunII_CDF_WJETS} (upper right) is the ratio 
${\sigma}_{W\ge n jets}$/${\sigma}_{W\ge n-1 jets}$ as a function of the jet 
multiplicity which gives a measure of the decrease in cross section with the addition 
of 1 jet. Being characterized by a reduced dependence on systematic uncertainties for both 
data and theory measurements this ratio is related to the magnitude of ${\alpha}_s$ (even if 
not giving a direct measure of it). For comparison, Run I results\cite{RunI_WJets} are
also reported in the same figure. The measured Run II over Run I cross 
section ratios are found to be in agreement with theoretical expectations.
In general, it can be concluded that a reasonable data to theory comparison is observed. 

Furthermore, the theory to data comparison was also investigated in some jet kinematic 
variable distributions: the differential cross section as a function of the jet $E_T$ and 
the dijet invariant mass ($M_{JJ}$) and angular separation (${\Delta R}_{JJ}$). 
Fig.\ref{RunII_CDF_WJETS} (lower left) shows the distributions for the leading jet $E_T$ in W 
$\ge$ 1 jets 
events, the second highest jet $E_T$ in W $\ge$ 2 jets events, and so on up to W $\ge$ 4 
jets events, respectively going from the top to the bottom plot. 
Fig.\ref{RunII_CDF_WJETS} (lower right) reports the $M_{JJ}$ distribution for the two 
leading jets in W $\ge$ 2 jets events. 
Data points include statistical errors, while the band represents the jet energy systematic 
uncertainty (the dominant one). The two lines are fits to the theoretical distributions 
calculated at different values for ${\mu}_{R/F}$. 
In general, a fair data to theory comparison is found within errors\footnote{
Theoretical uncertainties are characterized by a reduced dependence on ${\mu}_{R/F}$ 
with respect to the predictions on ${\sigma}_{W\ge n jets}$.
}
especially for the $M_{JJ}$ and ${\Delta R}_{JJ}$ distributions 
(which, at low values, are in particular sensitive to soft and collinear jet 
production) indicating that the LO calculation convoluted with the HERWIG parton shower 
approach, even if representing a partial higher order correction, can reproduce the data. 
However, the not 
negligible discrepancies between data and theory in the differential cross section 
at high ${E_T}^J$ can indicate some limitations of the parton radiation in the shower approach 
in properly describing the high multiplicity topologies\footnote
{
A major theoretical limitation in such kind of studies comes from the difficulty to 
proper merge matrix element calculations for different parton multiplicities avoiding 
double counting. For this reason only inclusive quantities are considered.
}.
\section{Underlying Event Studies}
The hard scattering process at hadron colliders is usually
accompanied by the so called ``enderlying event'' (UE) which, consisting of the 
contributions from beam-beam remnants, initial and final state radiation and (``semi-hard'') 
multiple parton interactions, has to be removed in order to achieve precise 
comparisons between jet measurements and pQCD predictions.
Consequently, an accurate modeling 
of the UE is important for all analyses involving jets in the final state, 
especially at low energies. 
At the same time our current understanding of this precess is limited as it involves both  
perturbative and non-perturbative QCD. 
It is clearly important to check how current QCD MC models describe the observed properties 
of the UE and, if possible, to device an optimal tuning so that to improve their 
fitting to data results.

\begin{figure}[!htb]
  \vspace{5.0cm}
  \includegraphics{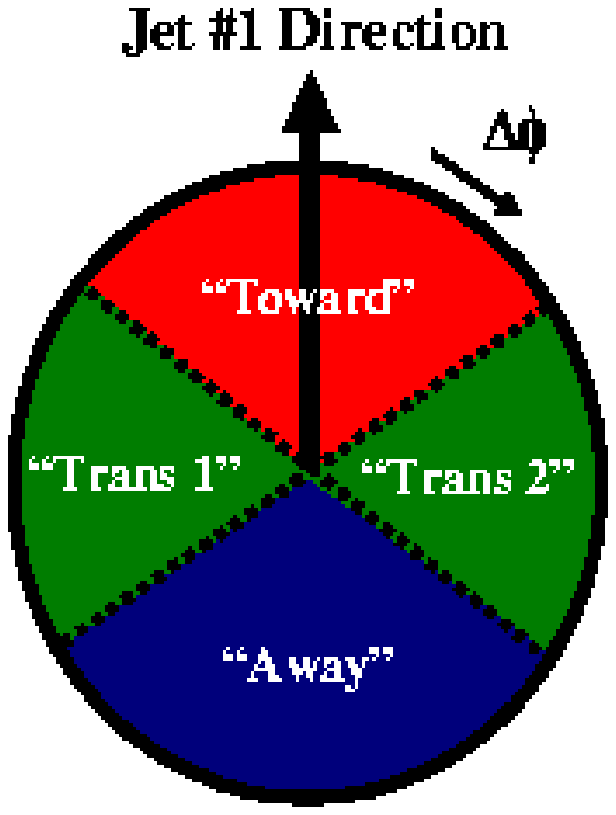}
  \includegraphics{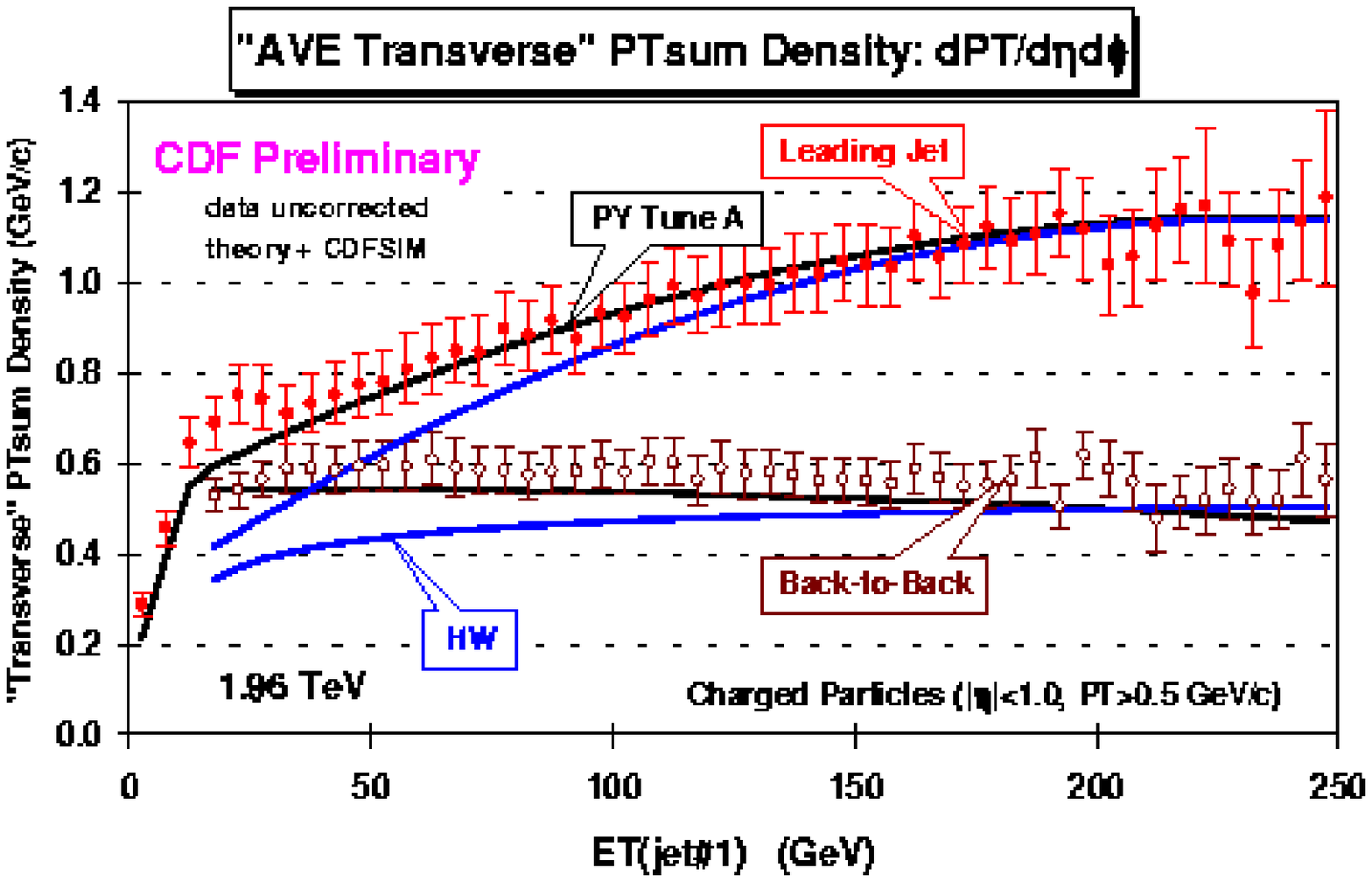}
  \caption{\it
    Left: The leading jet direction is used to define three regions in azimuthal angle
    (each spanning $120^o$) with different hadronic activities in the event. The 
``transverse''
    region is very sensitive to
    the UE contribution. Right: average charged particle $P_T$ sum density
    ($<$d$\Sigma{P_T}$/d$\eta$d$\phi$$>$) in the ``transverse'' region as a function of the
    leading jet $E_T$ (${E_T}^{J1}$) for ``leading jet'' and ``back-to-back'' events (as
    defined in the test) in Run II data, compared with predictions from HERWIG and PYTHIA 
(tuned
    on Run I data: ``Tune A'').
    \label{RunII_CDF_UE_stud} }
\end{figure}

Such kind of studies were performed in CDF during Run I\cite{CDF_I_UE}. 
The ``transverse'' region perpendicular to the leading jet of the event in the azimuthal 
angle (see fig.\ref{RunII_CDF_UE_stud}, left) is expected to be very sensitive to the UE 
and it was studied using charged particles
\footnote{
As the study of the UE mostly involves very low energy particles, only the charged 
particle component of the UE was considered because the CDF tracking system 
(immersed into a 1.4 T magnetic field) measures the momenta of low $P_T$ tracks more 
accurately than the calorimeter measures their energies.
}.
In general, it was observed that HERWIG\cite{herwig}, ISAJET\cite{isajet} and PYTHIA\cite{pythia} 
QCD MC models with their default parameters do not describe correctly all the properties of 
the ``transverse'' region when compared to minimum bias and jet trigger data.
In particular, it was found that a simple minimum bias modeling for the ``beam-beam remnants'' 
contribution to the UE (HERWIG, ISAJET) is not able to account for the observed 
charged particle multiplicity while, with the inclusion of multiple parton interactions, 
PYTHIA gives better results. It was actually found that PYTHIA, with an adequate tuning 
of its parameters in order to enhance this contribution, is able to fit data very 
well\cite{PythTuneA}. 

Similar studies have been performed in CDF using Run II data at 
$\sqrt{s}$ = 1.96 TeV. 
As in Run I, the topological structure of $p\bar p$ collisions was considered in order to 
make a phenomenological study of the UE in minimum bias and jet trigger data samples to be 
compared to HERWIG and PYTHIA QCD MC predictions. 
In the study reported here, the direction of the leading calorimeter jet (JetClu R = 0.7, 
$|\eta|$ $<$ 2) was used to isolate regions of ${\eta}$-${\phi}$ space that are sensitive 
to the UE (fig.\ref{RunII_CDF_UE_stud}, left) and observables related to charged 
particles (reconstructed by the central tracking system with 
$P_T$ $>$ 0.5 GeV/c and $|\eta|$ $<$ 1) were considered. 
Two classes of events were defined according to the jet topology: ``leading jet'' events, 
with no restriction applied on the second highest $E_T$ jet and ``back-to-back'' events, 
where the two jets were required to be nearly back-to-back 
(${\Delta}{\phi}_{12}$ $>$ $150^o$, ${E_T}^{J2}$/${E_T}^{J1}$ $>$ 0.8). 
This classification was introduced in order to select a subsample (``back-to-back'' configuration) 
where hard initial and final state radiation are suppressed to increase the 
sensitivity of the ``transverse'' region to the ``beam-beam remnants'' and multiple parton 
scattering component of the UE.    
Fig.\ref{RunII_CDF_UE_stud} (right) shows the average charged particle $P_T$ sum density 
($<$d$\Sigma{P_T}$/d$\eta$d$\phi$$>$) in the ``transverse'' region as a function of the leading jet 
$E_T$ (${E_T}^{J1}$) in ``leading jet'' and ``back-to-back'' events. Also shown are the 
predictions from HERWIG and PYTHIA after a full detector simulation.
The charged track activity is different for the two configurations: for the ``leading jet'' 
case the density rises with increasing ${E_T}^{J1}$, while for the ``back-to-back'' one it 
slightly falls  with increasing ${E_T}^{J1}$. The rise for ``leading jet'' events 
is attributed to hard initial and final state radiation which has been suppressed 
in ``back-to-back'' events whose opposite trend with increasing ${E_T}^{J1}$ might be due 
to a ``saturation'' of the multiple parton interaction at small impact parameter.
Such effect is expected to be included in PYTHIA (with multiple parton interactions included) 
but not in HERWIG (without multiple parton interactions). 
Indeed, in fig.\ref{RunII_CDF_UE_stud} (right) we see how PYTHIA tuned on Run I data 
(``Tune A'', version 6.206) fits very well the data for both configurations, while HERWIG looks 
working only at higher ${E_T}^{J1}$.
Similar results are observed considering the average charged particle number density 
($<$dN/d$\eta$d$\phi$$>$) as well as in additional studies involving, for instance, the 
${\Delta}{\phi}$ dependence of these densities (d$\Sigma{P_T}$/d$\eta$d$\phi$ and 
dN/d$\eta$d$\phi$) relative to the direction of the leading jet.

\section{Conclusions}
A very exciting and important QCD physics program is ongoing at the Tevatron $p\bar p$ 
collider where the increase in center-of-mass energy from 1.8 to 
1.96 TeV and the higher statistics of Run II data samples are expected to extend Run I 
results at high ${E_T}^J$.

Some preliminary Run II results have been reported in this contribution. 
The measured inclusive jet and dijet mass cross sections are in 
reasonable agreement with NLO pQCD (EKS, JETRAD) within the uncertainty on jet energy 
scale in the data and on gluon PDF at high $x$ in the theoretical predictions. 
The production of W + jets events is fairly described by an ``enhanced'' LO QCD 
Monte Carlo (ALPGEN + HERWIG). Data to theory comparison is limited by the systematic 
error on the jet energy scale and by the ${\mu}_{R/F}$ dependence of LO calculations.
The underlying event is well described by the model implemented in PYTHIA using 
parameters tuned on Run I data. 
\section{Acknowledgements}
I'm very grateful to the Conference Organizers, G. Bellettini, G. Chiarelli and M. Greco, 
for their warm hospitality in the unique atmosphere of La Thuile. \\
I would like also to acknowledge the QCD Group members of the CDF and D$\0$ 
Collaborations for their work in achieving the results shown in this presentation. 
Many thanks to Mario Martinez for his precious comments leading to the final 
version of the manuscript.
\end{document}